\documentclass[10pt,twocolumn]{article}
\usepackage{times}
\usepackage{float}
\usepackage[noend]{distribalgo}
\usepackage{algorithm}
\usepackage{fullpage}
\usepackage{subfig}
\usepackage{multirow}
\usepackage{graphicx}
\usepackage{bsymb}
\usepackage{comment}
\usepackage{color}

\newcommand{\mysection}[1]{
	\section{#1}
}
\newcommand{\mysubsection}[1]{
	\subsection{#1}
}
\newcommand{\mysubsubsection}[1]{
	\subsubsection{#1}
}
\newcommand{\myparagraph}[1]{
	\paragraph{#1}
}

\newlength{\myitemsep}
\newlength{\mytopsep}
\newlength{\myparsep}
\newlength{\myparskip}
\newlength{\myleftmargin}
\newcommand{\init}[0]{\textbf{initially}}
\newcommand{\var}[0]{\textbf{var}}

\newcommand{\intransition}[0]{\textbf{internal transition}}
\newcommand{\outtransition}[0]{\textbf{interface transition}}
\newcommand{\precond}[0]{\textbf{precondition}}
\newcommand{\action}[0]{\textbf{action}}

\newcommand{\client}[0]{\mathit{clt}}
\newcommand{\replica}[0]{\mathit{replica}}
\newcommand{\proposals}[0]{\mathit{proposals_{replica}}}

\newcommand{\inputs}[0]{\mathit{inputs}_\nu}
\newcommand{\outputs}[0]{\mathit{outputs}_\nu}

\newcommand{\decisions}[0]{\mathit{decisions}}
\newcommand{\learned}[0]{\mathit{learned}_{replica}}
\newcommand{\curstate}[0]{\mathit{appState_{replica}}}
\newcommand{\shadowstate}[0]{\mathit{shadowState_{replica}}}
\newcommand{\shadowversion}[0]{\mathit{shadowVersion_{replica}}}
\newcommand{\state}[0]{\mathit{appState}}
\newcommand{\newState}[0]{\mathit{newState}}

\newcommand{\version}[0]{\mathit{version_{replica}}}

\newcommand{\op}[0]{\mathit{op}}

\newcommand{\result}[0]{\mathit{result}}
\newcommand{\received}[0]{\mathit{responded_{\client}}}
\newcommand{\invoked}[0]{\mathit{invoked_{\client}}}

\newcommand{\cert}[0]{\mathit{cert}}
\newcommand{\coord}[0]{\mathit{proseq}}

\newcommand{\cmd}[0]{\mathit{cmd}}

\newcommand{\bevId}[0]{\mathit{rid}}
\newcommand{\progind}[0]{\mathit{prog}}
\newcommand{\certBev}[0]{\mathit{rid_{cert}}}

\newcommand{\certified}[0]{\mathit{certifics}_\nu}

\newcommand{\snapshots}[0]{\mathit{snapshots}_\nu}
\newcommand{\progress}[0]{\mathit{progress_{cert}}}

\newcommand{\primary}[0]{\mathit{isSeq_{cert}}}
\newcommand{\certId}[0]{\mathit{cert}}
\newcommand{\slot}[0]{\mathit{slot}}

\newcommand{\round}{round}
\newcommand{\Round}{Round}

\newcommand{\sequencer}[0]{sequencer}

\newcommand{\Sequencer}[0]{Sequencer}

\newcommand{\nextState}[0]{\mathit{nextState}}

\newcommand{\multiconsensus}[0]{Multi-Consensus}

\newcommand{\observedecision}[0]{observeDecision}

\newenvironment{definition}[1]{\begin{trivlist}
\item[\hskip \labelsep {\bfseries #1:}]}{\end{trivlist}}

\floatname{algorithm}{Specification}

\title{Vive La Diff\'erence:\\
Paxos vs. Viewstamped Replication vs. Zab}
\author{Robbert van Renesse, Nicolas Schiper, Fred B. Schneider\\
Department of Computer Science, Cornell University}

\date{}

\begin{document}

\maketitle

\begin{abstract}

Paxos, Viewstamped Replication, and Zab are replication protocols
that ensure high-availability in asynchronous environments with crash
failures. Various claims have been made about similarities and
differences between these protocols. But how does one determine
whether two protocols are the same, and if not, how significant the
differences are?

We propose to address these questions using refinement mappings,
where protocols are expressed as succinct specifications that are
progressively refined to executable implementations.  Doing so enables
a principled understanding of the correctness of the different
design decisions that went into implementing the various protocols.
Additionally, it allowed us to identify key differences
that have a significant impact on performance.

\end{abstract}

\mysection{Introduction}

A protocol expressed in terms of a state transition specification $\Sigma$
refines another specification $\Sigma'$ if there exists a mapping of the
state space of $\Sigma$ to the state space of $\Sigma'$ and
each state transition in $\Sigma$
can be mapped
to a state transition in $\Sigma'$ or to a no-op.
This mapping between specifications is called
\emph{refinement}~\cite{Lam83} or \emph{backward
simulation}~\cite{LynchV96}.
If two protocols refine one another then we might argue that they are alike.
But if they don't, how does one qualify
the similarities and differences between two protocols?

We became interested in this question while comparing three replication
protocols for high availability in asynchronous environments with
crash failures:

\begin{itemize}
\item \emph{Paxos}~\cite{L98} is a state
machine replication protocol~\cite{Lam78,S90}.
We consider
a version of Paxos that uses the multi-decree Synod consensus algorithm
described in~\cite{L98}, sometimes called Multi-Paxos.
Many implementations have been deployed,
including in Google's Chubby service~\cite{Bur06,CGR07},
in Microsoft's Autopilot service~\cite{Isa07} (used by Bing), and in the
popular Ceph distributed file system~\cite{WBM06},
with interfaces now part of the standard Linux kernel.
\item \emph{Viewstamped Replication} (VSR)~\cite{OL88,liskov12vr} is a
replication protocol originally targeted at replicating participants
in a Two-Phase Commit (2PC)~\cite{LS76} protocol.
VSR has also been used in the implementation of the
Harp File System~\cite{LGGJSW91};
\item Zab~\cite{JRS11} (ZooKeeper Atomic Broadcast) is a replication
protocol used for the popular \emph{ZooKeeper}~\cite{HKJR10} configuration
service.  ZooKeeper has been in active use at Yahoo! and
is now a popular open source product distributed by Apache.
\end{itemize}

\noindent
Many claims have been made about the similarities
and differences of these protocols.
For example, citations~\cite{Lampson93,Cac09}
claim that Paxos and Viewstamped Replication are ``the same algorithm
independently invented,'' ``equivalent,'' or that ``the view management
protocols seem to be equivalent''~\cite{L98}.

\begin{figure}
\centering
\includegraphics[scale=0.5]{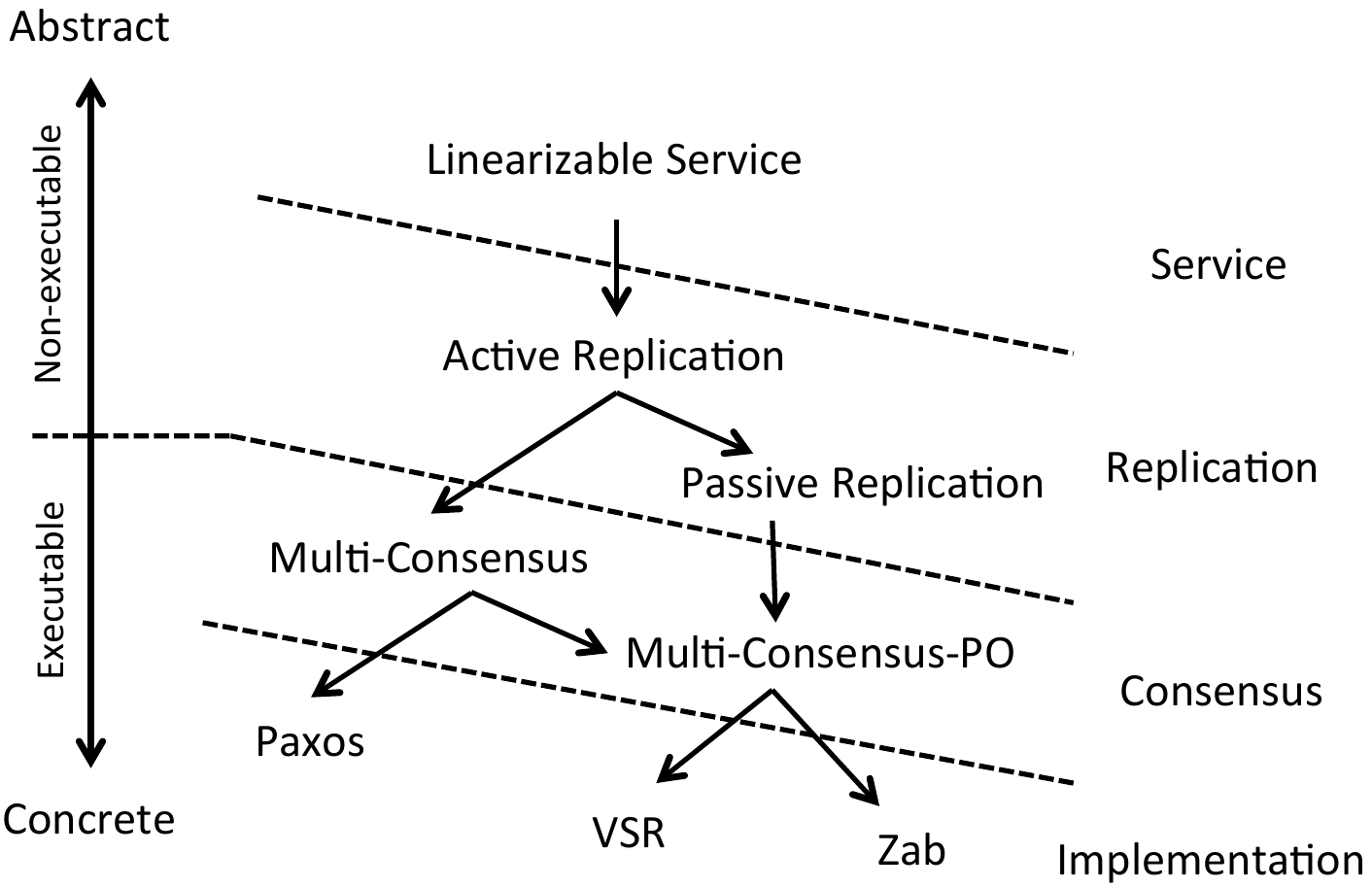}
\caption{{\small An ordering of refinement mappings and informal levels of abstraction.}}
\label{fig:lattice}
\end{figure}

In this paper, we approach the question of similarities and differences
between these protocols using refinements, as illustrated in
Fig.~\ref{fig:lattice}.
Refinement mappings induce an ordering relation on specifications, and
the figure shows a Hasse diagram of a set of eight specifications of
interest ordered by refinement.
In this figure, we write $\Sigma' \rightarrow \Sigma$ if $\Sigma$ refines
$\Sigma'$, that is, if there exists a
refinement mapping of $\Sigma$ to $\Sigma'$.

At the same time, we have indicated informal levels of abstraction in this
figure, ranging from a highly abstract specification of
a linearizable service~\cite{HW90} to concrete, executable specifications.
Active and passive replication are common
approaches to replicate a service and ensure that behaviors are still
linearizable.
{\multiconsensus} protocols use a form of rounds in order to refine active and passive replication.
Finally, we obtain protocols such as Paxos, VSR, and Zab.

Each refinement corresponds to a \emph{design decision}, and
as can be seen from the figure it is possible to arrive at the same
specification following different paths of such design decisions.  There is a
qualitative difference between refinements that cross abstraction
boundaries and those that do not.
When crossing an abstraction boundary, a refinement takes an abstract
concept and replaces it with a more concrete one.  For example, it may
take an abstract \emph{decision} and replace it by a \emph{majority of votes}.
When staying within the same abstraction, a refinement \emph{restricts}
behaviors.  For example, one specification might decide commands out of
order while a more restricted specification might decide them in order.

Using these refinements, we can identify and analyze
commonalities and differences between the replication protocols.
They also enable consideration of new variants of these protocols
and what conditions they must satisfy to be correct refinements
of a linearizable service.
Other papers have used refinements to specify a replication protocol~\cite{Lampson01,Lamport11}.
To the best of our knowledge, this paper is the first to employ refinement for
\emph{comparing} different replication protocols.

This paper is organized as follows:
Section~\ref{sec:replication} introduces state transition
specifications of linearizable replication as well as active and passive replication.
Section~\ref{sec:bevs} presents \multiconsensus, a canonical protocol
that generalizes Paxos, VSR, and Zab
and forms a basis for comparison. 
\emph{Progress indicators}, a new class of invariants,
enable constructive reasoning about why and how these protocols work.
We show how passive replication
protocols refine \multiconsensus\ by adding \emph{prefix order} (\emph{aka} \emph{primary order}), rendering \multiconsensus-PO.
Section~\ref{sec:implementation} presents implementation
details of Paxos, VSR, and Zab that constitute the final
refinement steps to executable protocols.
Section~\ref{sec:discuss} discusses the implications of
identified differences on performance.
Section~\ref{sec:history} gives a short overview of the
history of concepts used in these replication protocols, and
Section~\ref{sec:conclusion} is a conclusion.

\mysection{Masking Failures}
\label{sec:replication}

To improve the availability of a service, a common technique is
to replicate it onto a set of servers.
A \emph{consistency criterion} defines expected responses
to clients for concurrent operations.  Ideally, the replication
protocol ensures \emph{linearizability}~\cite{HW90}---the execution of
concurrent client operations is equivalent to a sequential execution,
where each operation is atomically performed at some point in time between its
invocation and response.

\mysubsection{Specification}
We characterize linearizability by giving a state transition
specification (see Specification~\ref{alg:replsvc}).  A specification
defines states and gives legal transitions between states.
A state is defined by a collection of variables and their current values.
Transitions can involve
parameters (listed in parentheses) that are bound within the defined scope.
A transition definition gives a precondition and an action.
If the precondition holds in a given state,
then the transition is \emph{enabled} in that state.
The action relates the state after the transition to
the state before.
A transition is performed indivisibly starting in a state satisfying
the precondition.  No two transitions are performed concurrently, and
if multiple transitions are enabled simultaneously, then the choice of
which transition to perform is unspecified.

There are \emph{interface variables} and \emph{internal variables}.
Interface variables are subscripted with the location of the variable, which
is either a process name or the network, $\nu$.
Internal variables have no subscripts, and their value will be determined
by a function on the state of the underlying implementation.
Specification Linearizable Service has the following variables:

\begin{itemize}
\item $\inputs$: a set that contains $(\client, \op)$ messages
sent by process $\client$.  Here $\op$ is an operation invoked by $\client$;
\item $\outputs$: a set of $(\client, \op, \result)$ messages sent
by the service, containing the results of client operations that have been
executed;
\item $\state$: an internal variable containing the state of the application;
\item $\invoked$: the set of operations invoked by process $\client$.  This is a
variable maintained by $\client$ itself;
\item $\received$: the set of completed operations, also maintained by $\client$.
\end{itemize}

\begin{algorithm}[t]
\caption{\label{alg:replsvc} Linearizable Service}
\footnotesize{
\begin{distribalgo}[0]
\label{spec_object}

\STATE \var ~ $\inputs, \outputs, \state, \invoked, \received$

\BLANK

\INDENT{\init: $\state = \bot \land \inputs = \outputs = \emptyset ~\land$}
\STATE $\forall \client : \invoked = \received = \emptyset$
\ENDINDENT

\BLANK

\INDENT{\outtransition\ \texttt{invoke}($\client, \op$):}
\INDENT{\precond:}
  \STATE $\op \not\in \invoked$
\ENDINDENT
\INDENT{\action:}
  \STATE $\invoked := \invoked \cup \lbrace op \rbrace$
  \STATE $\inputs := \inputs \cup \lbrace (\client, op) \rbrace$
\ENDINDENT
\ENDINDENT

\BLANK

\INDENT{\intransition\ $\texttt{execute}(\client, op, \result, \newState)$:}
\INDENT{\precond:}
  \STATE $(\client, op) \in \inputs
~\land$
  \STATE $(\result, \newState) = \nextState(\state, (\client, op))$
\ENDINDENT
\INDENT{\action:}
  \STATE $\state := \newState$
  \STATE $\outputs := \outputs \cup \lbrace ((\client, op), \result) \rbrace$
\ENDINDENT
\ENDINDENT

\BLANK

\INDENT{\outtransition\ \texttt{response}($\client, \op, \result$):}
\INDENT{\precond:}
  \STATE $((\client, \op), \result) \in \outputs \land \op \not\in \received$
\ENDINDENT
\INDENT{\action:}
  \STATE $\received := \received \cup \lbrace \op \rbrace$
\ENDINDENT
\ENDINDENT

\end{distribalgo}
}
\end{algorithm}

\noindent
Similarly, there are \emph{interface transitions} and
\emph{internal transitions}.
Interface transitions model interactions with the environment, which
consists of a collection of processes connected by a network.
An interface transition is performed by
the process that is identified by the first parameter to the transition.
Interface transitions are not allowed to access internal variables.
Internal transitions are performed by the service, and we will have
to demonstrate how this is done by implementing those transitions.
The transitions of Specification Linearizable Service are:

\begin{itemize}
\item Interface transition $\texttt{invoke}(\client, \textit{op})$ is
performed when $\client$ invokes
operation \textit{op}.  Each operation is uniquely identified
and can be invoked at most once by a client (enforced by the precondition).
Adding $(\client, \op)$ to $\inputs$
models $\client$ sending a message containing $\op$ to the service.
The client maintains what operations it has invoked in $\invoked$;
\item Transition
$\texttt{execute}(\client, \textit{op}, \result, \textit{newState})$
is an internal transition that is performed when the replicated service executes \textit{op} for
client $\client$.
The application-dependent deterministic function
$\nextState$ relates an application state and an operation from a client
to a new application state and a result.
Adding $((\client, \op), \result)$ to $\outputs$ models the service
sending a response to $\client$.
\item Interface transition
$\texttt{response}(\client, \textit{op}, \result)$
is performed when $\client$ receives the response.
The client keeps track of which operations have completed
in $\received$ to prevent this transition being performed more than
once per operation.
\end{itemize}

\noindent
From Specification Linearizable Service
it is clear that it is not possible for a client
to receive a response to an operation before it has been invoked and executed.
However, the specification does allow each client operation to be executed
an unbounded number of times.
In an implementation, multiple execution could happen if the response to the
client operation got lost by the network and the client retransmits its operation
to the service.
The client will only learn about at most one of these executions.
In practice, replicated services will try to reduce or
eliminate the probability of a client operation being executed more than
once by keeping state about which operations have been executed.
For example,
a service could keep track of all its clients and
eliminate duplicate operations using sequence numbers on client operations.  In all the specifications that follow,
we omit the logic to avoid operations from being executed multiple times to simplify the presentation.

We make the following assumptions about interface transitions:

\begin{itemize}
\item \textbf{Crash Failures}:
A process follows its specification until it fails by crashing.
Thereafter, it executes no transitions.
Processes that never crash are called \emph{correct}.
A process that ``shuts down'' and later recovers using state from stable storage
is considered correct albeit, temporarily slow.
Processes are assumed to fail independently.
\item \textbf{Failure Threshold}:
There is a bound $f$ on the maximum number of replica processes that may crash; the number of client processes that fail is unbounded.
\item \textbf{Fairness}:
Except for interface transitions at a crashed process, a transition that
becomes continuously enabled is eventually executed.
\item \textbf{Asynchrony}:
There is no bound on the time before a continuously enabled transition
is executed.
\end{itemize}


We will use refinement only to show that the design decisions that
we introduce are safe.  The intermediate specifications that we will
produce will not necessarily guarantee liveness.  It is important
that the final executable implementations support liveness properties
such as ``an operation issued by a correct client is eventually
executed,'' but it is not necessary for our purposes that intermediate
specifications have such liveness properties.\footnote{State transition specifications may also
include supplementary liveness conditions.  If so, a specification $\Sigma$ that refines a specification $\Sigma'$ preserves both the \emph{safety} and \emph{liveness} properties of $\Sigma'$.}

\mysubsection{Active and Passive Replication}\label{sec:actpass}

Specification~\ref{alg:replsvc} has internal variables and transitions
that have to be implemented.
There are two well-known approaches to replication:

\begin{itemize}
\item With \emph{active replication}, also known as \emph{state
machine replication}~\cite{Lam78,S90},
each replica implements a
deterministic state machine.
All replicas process the same operations in the same order.
\item With \emph{passive replication}, also known as \emph{primary
backup}~\cite{AD76}, a primary replica runs a deterministic
state machine, while backups only store states.
The primary computes a sequence of new application states by processing
operations and forwards these states to each backup in order of generation.
\end{itemize}

\begin{figure*}
  \centering
  \subfloat[Active replication]{
    \label{fig:active_repl}
    \includegraphics[scale=.6]{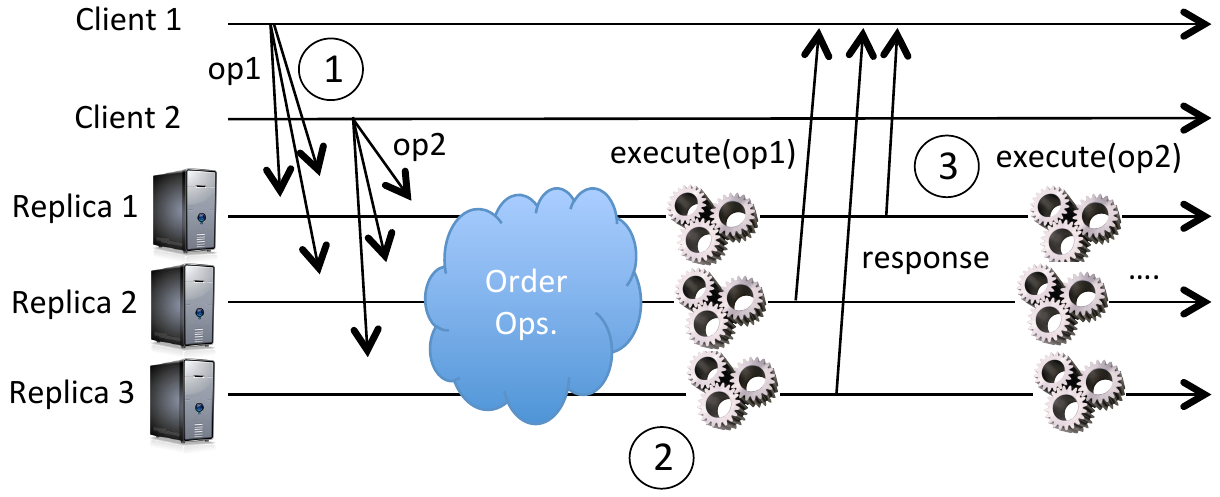}
  }
  \subfloat[Passive replication]{
    \label{fig:passive_repl}
    \includegraphics[scale=.6]{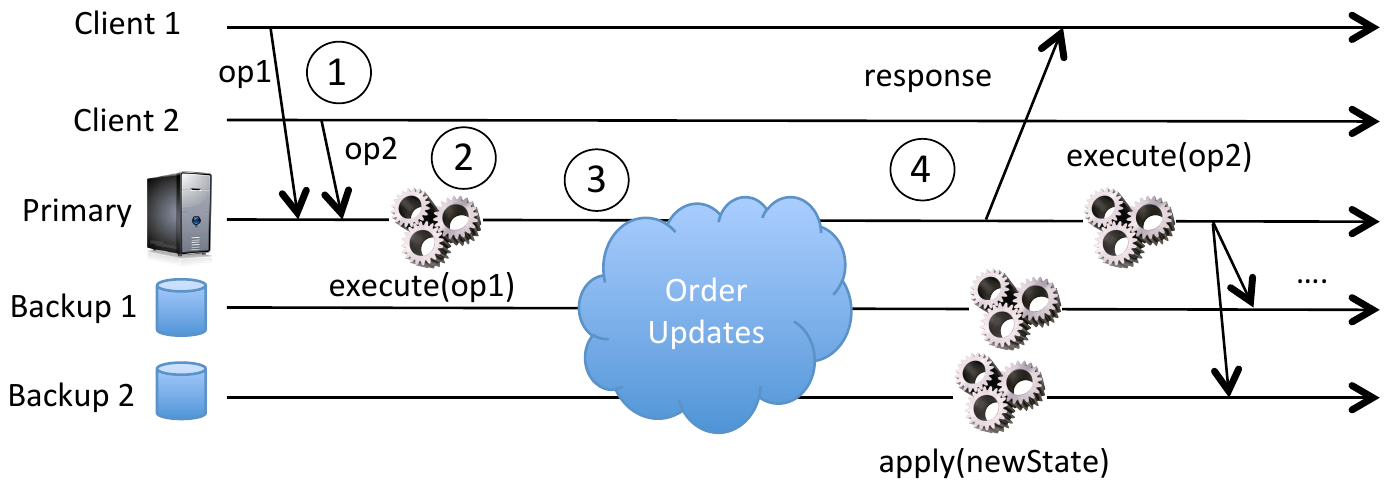}
  }
  \caption{{\small A failure-free execution of active and passive replication.}}
  \label{fig:active_passive_repl}
\end{figure*}

\noindent
Fig.~\ref{fig:active_repl} illustrates a failure-free execution
of a service implemented using active replication:
\begin{enumerate}
\item Clients submit operations to the service (\texttt{op1} and \texttt{op2} in Fig.~\ref{fig:active_repl}).
\item Replicas, starting out in the same state, execute
received client operations in the same order.
\item Replicas send responses to the clients.  Clients ignore
all but the first response they receive.
\end{enumerate}

\noindent
The tricky part of active
replication is ensuring that replicas execute operations in
the same order, despite replica failures, message loss, and unpredictable
delivery and processing delays. A fault-tolerant
\emph{consensus} protocol~\cite{PSL80} is typically employed
for replicas to agree on the $i^\textit{\footnotesize th}$ operation
for each index $i$ in a sequence of operations.
Specifically, each replica proposes an
operation that was received from one of the clients in instance $i$
of the consensus protocol.
Only one of the proposed operations can be decided.
The service remains available as long as each instance of consensus 
eventually terminates.


Fig.~\ref{fig:passive_repl} depicts a
failure-free execution of passive replication:
\begin{enumerate}
\item Clients submit operations only to the primary.
\item The primary orders the operations and computes new
states and responses.
\item The primary forwards the new states (so-called \emph{state updates})
to each backup in the order generated.
\item The primary sends the response of an
operation only after the corresponding state update has been successfully
\emph{decided} (this is made precise in Section~\ref{sec:bevs}).
\end{enumerate}


\noindent
Because two primaries may be competing to have their state updates applied at the backups, it is important that replicas apply a state update $u$
on the same state used by the primary to compute $u$.  
This is sometimes called the \emph{prefix order} or \emph{primary order} property~\cite{JRS11,JS13}.

For example, consider a replicated integer variable with initial
value~3.
One client wants to increment the variable, while the other wants
to double it.
One primary receives both operations and submits state updates~4
followed by~8.  Another primary receives the operations in the
opposite order and submits updates~6 followed by~7.
Without prefix ordering, it may happen that the decided states
are~4 followed by~7, not corresponding to any sequential history of the
two operations.

VSR and Zab employ passive replication; Paxos employs active
replication.
However, it is possible to implement one style on the other.
The Harp file system~\cite{LGGJSW91}, for example,
uses VSR to implement a replicated message queue
containing client operations---the Harp primary proposes state
updates that backups apply to the state of the message queue.
Replicas, running deterministic NFS state machines, then execute
NFS operations in queue order.  In other words,
Harp uses an active replication protocol built using a message queue
that is passively replicated using VSR.


\mysubsection{Refinement}\label{sec:ap-formal}

Below, we present a refinement of a linearizable service
(Specification~\ref{alg:replsvc})
using the active replication approach.  We then further refine active
replication to obtain passive replication.  The
refinement of a linearizable service to passive replication follows
transitively.

\subsubsection{Active Replication}
We omit interface transitions $\texttt{invoke}(\client, \op)$
and $\texttt{response}(\client, \op, \result)$,
which are the same as in Specification~\ref{alg:replsvc}.  Hereafter,
a command $\cmd$ denotes a tuple $(\client, \op)$.

Specification~\ref{alg:active}
uses a sequence of \emph{slots}.
A replica executes transition $\texttt{propose}(\replica, \slot, \cmd)$ to
propose a command $\cmd$ for the given $\slot$.
We call the command a \emph{proposal}.
Transition
$\texttt{decide}(\slot, \cmd)$ guarantees that at most one proposal
is decided for each slot.  
Transition $\texttt{learn}(\replica, \slot)$ models a replica learning a decision and assigning the decision to the corresponding slot of the $\learned$ array.  
Replicas update their state by executing a learned operation in
increasing order of slot number with transition $\texttt{update}(\replica, \cmd, \textit{res}, \textit{newState})$.  The slot of the next operation
to execute is denoted by $\version$.




\begin{figure}
\begin{center}
 \begin{algorithm}[H]

\caption{\label{alg:active} Specification Active Replication}

\footnotesize{
\begin{distribalgo}[0]
 
\INDENT{\var ~ $\proposals[1...], \decisions[1...], \learned[1...]$}
\STATE $~~~\curstate, \version, \inputs, \outputs$
\ENDINDENT

\BLANK

\INDENT{\init:}
\STATE $\forall s \in \nat^{+}: \decisions[s] = \bot ~\land$
\STATE $\forall \mathit{replica} :$
\STATE $~~~\curstate = \bot \land \version = 1 ~\land$
\STATE $~~~\forall s \in \nat^{+}:$
   \STATE $~~~~~\proposals[s] = \emptyset \land \learned[s] = \bot$
\ENDINDENT

\BLANK

\INDENT{\outtransition\ $\texttt{propose}(\replica, \slot, \cmd)$:}
\INDENT{\precond:}
  \STATE $\cmd \in \inputs \land \learned[\slot] = \bot$
\ENDINDENT
\INDENT{\action:}
  \STATE $\proposals[\slot] := \proposals[\slot] \cup \{ \cmd \}$
\ENDINDENT
\ENDINDENT

\BLANK

\INDENT{\intransition\ $\texttt{decide}(\slot, \cmd)$:}
\INDENT{\precond:}
  \STATE $\decisions[\slot] = \bot ~\land ~\exists r: \cmd \in \textit{proposals}_r[\slot]$
\ENDINDENT
\INDENT{\action:}
  \STATE $\decisions[\slot] := \cmd$
\ENDINDENT
\ENDINDENT

\BLANK

\INDENT{\intransition\ $\texttt{learn}(\replica, \slot)$:}
\INDENT{\precond:}
  \STATE $\learned[\slot] = \bot \land \decisions[\slot] \ne \bot$
\ENDINDENT
\INDENT{\action:}
  \STATE $\learned[\slot] := \decisions[\slot]$
\ENDINDENT
\ENDINDENT

\BLANK

\INDENT{\outtransition\ $\texttt{update}(\replica, \cmd, \textit{res}, \newState)$:}
\INDENT{\precond:}
  \STATE $\cmd = \learned[\version] \land \cmd \ne \bot ~\land$ \\
  \STATE $(\textit{res}, \newState) = \nextState(\curstate, \cmd)$
\ENDINDENT
\INDENT{\action:}
  \STATE $\outputs := \outputs \cup \lbrace (\cmd, \textit{res}) \rbrace$
  \STATE $\curstate:= \newState$
  \STATE $\version := \version + 1$
\ENDINDENT
\ENDINDENT
\end{distribalgo}
}
\end{algorithm}
\end{center}
\end{figure}

\begin{figure}
\begin{center}
\begin{algorithm}[H]
\caption{\label{alg:passive} Specification Passive Replication}

\footnotesize{
\begin{distribalgo}[0]

\INDENT{\var ~ $\proposals[1...], \decisions[1...], \learned[1...]$}
\STATE $~~~\curstate, \version,\inputs, \outputs,$
\STATE $~~~\shadowstate, \shadowversion$
\ENDINDENT

\BLANK

\INDENT{\init:}
\STATE $\forall s \in \nat^{+}: \decisions[s] = \bot$
\STATE $\forall \mathit{replica} :$
\STATE $~~~\curstate = \bot \land \version = 1 ~\land$
\STATE $~~~\shadowstate = \bot \land \shadowversion = 1 ~\land$
\STATE $~~~\forall s \in \nat^{+}:$
   \STATE $~~~~~ \proposals[s] = \emptyset \land \learned[s] = \bot$
\ENDINDENT

\BLANK

\INDENT{\outtransition\ $\texttt{propose}(\replica, \slot, \cmd,$\\ 
								\hspace{2in} $\textit{res}, \textit{newState})$:}
\INDENT{\precond:}
  \STATE $\cmd \in \inputs \land \slot = \shadowversion ~\land$
  \STATE $\learned[\slot] = \bot ~\land$
  \STATE $(\textit{res}, \textit{newState}) = \nextState(\shadowstate, \cmd)$
\ENDINDENT
\INDENT{\action:}
  \STATE $\proposals[\slot] := \proposals[\slot] ~\cup$
  \STATE $~~~~~~~~~~~~~~\{ (\shadowstate, (\cmd, \textit{res}, \textit{newState})) \}$
  \STATE $\shadowstate := \textit{newState}$
  \STATE $\shadowversion := \slot + 1$
\ENDINDENT
\ENDINDENT

\BLANK

\INDENT{\intransition\ $\texttt{decide}(\slot, \cmd, \textit{res}, \textit{newState})$:}
\INDENT{\precond:}
  \STATE $\decisions[\slot] = \bot ~\land$
  \STATE $\exists r, s: (s, (\cmd, \textit{res}, \textit{newState})) \in \textit{proposals}_r[\slot]~\land$
  \STATE $~~~~~~~~~~(\slot > 1 \Rightarrow \decisions[\slot-1] = (-, -, s))$
\ENDINDENT
\INDENT{\action:}
  \STATE $\decisions[\slot] := (\cmd, \textit{res}, \textit{newState})$
\ENDINDENT
\ENDINDENT

\BLANK

\INDENT{\intransition\ $\texttt{learn}(\replica, \slot)$:}
\INDENT{\precond:}
  \STATE $\learned[\slot] = \bot \land \decisions[\slot] \ne \bot$
\ENDINDENT
\INDENT{\action:}
  \STATE $\learned[\slot] := \decisions[\slot]$
\ENDINDENT
\ENDINDENT

\BLANK

\INDENT{\outtransition\ $\texttt{update}(\replica, \cmd, \textit{res}, \textit{newState})$:}
\INDENT{\precond:}
  \STATE $(\cmd, \textit{res}, \textit{newState}) = \learned[\version] $
\ENDINDENT
\INDENT{\action:}
  \STATE $\outputs := \outputs \cup \lbrace (\cmd, \textit{res}) \rbrace$
  \STATE $\curstate := \textit{newState}$
  \STATE $\version := \version + 1$
\ENDINDENT
\ENDINDENT

\BLANK

\INDENT{\outtransition\ $\texttt{resetShadow}(\replica, \textit{version}, \textit{state})$:}
\INDENT{\precond:}
  \STATE $\textit{version} \ge \version ~\land$
  \STATE $(\textit{version} = \version \Rightarrow \textit{state} = \curstate)$
\ENDINDENT
\INDENT{\action:}
   \STATE $\shadowstate := \textit{state}$
   \STATE $\shadowversion := \textit{version}$
\ENDINDENT
\ENDINDENT

\end{distribalgo}
}
\end{algorithm}
\end{center}
\end{figure}


Note that $\texttt{propose}(\replica, \slot, \cmd)$ requires that 
$\replica$ has not yet learned a decision in $\slot$.  While not
a necessary requirement for safety, proposing a command for
a slot that is known to be decided is wasted effort.
It would make sense to require that replicas propose in the first
$\slot$ for which both $\proposals[\slot] = \emptyset$ and
$\learned[\slot] = \bot$.
We do not require this at this level of specification to
simplify the refinement mapping between active and passive replication.

To show that active replication refines
Specification~\ref{alg:replsvc}, we first show how
the internal state of Specification~\ref{alg:replsvc} is derived
from the state of Specification~\ref{alg:active}.
The only internal state in Specification~\ref{alg:replsvc} is the
$\state$ variable.
For our refinement mapping, its value is
the application state maintained by the replica (or one of the replicas)
with the highest version number.

To complete the refinement mapping we also have to show how transitions
of active replication map onto enabled transitions of
Specification~\ref{alg:replsvc}, or onto \emph{stutters} (no-ops with
respect to Specification~\ref{alg:replsvc}).
The \texttt{propose}, \texttt{decide}, and \texttt{learn} transitions are always stutters
because they do not update $\curstate$ of any replica.
An $\texttt{update}(\replica, \cmd, \textit{res}, \newState)$
transition corresponds to
$\texttt{execute}(\client, op, \textit{res}, \newState)$, where $\cmd = (\client, op)$,
in Specification~\ref{alg:replsvc}
if $\replica$ is the first replica to apply $\op$ and thus leading
to $\state$ being updated.
Transition $\texttt{update}$ is a stutter if the replica is not the
first replica to apply the update.

\subsubsection{Passive Replication}

Passive replication (Specification~\ref{alg:passive}) also uses
slots, and proposals are tuples
$(\textit{oldState}, (\cmd, \textit{res}, \textit{newState}))$
consisting of the state prior to executing a command, a command, the
output of executing the command, and a new state that results
from applying the command.
In an actual implementation, the old state and new state would respectively be 
represented by an identifier and a state update rather than the entire value of
the state.

Any replica can act as primary. 
Primaries act \emph{speculatively}, computing a sequence of states
before they are decided.
Because of this, primaries have to maintain a separate version of
the application state.  We call this the \emph{shadow state}.
Primaries may propose to apply different state updates for the same slot.

Transition
$\texttt{propose}(\replica, \slot, \cmd, \textit{res}, \textit{newState})$
is performed when a primary $\replica$
proposes applying $\cmd$ to $\shadowstate$ in a
certain $\slot$, resulting in output \textit{res}.
State $\shadowstate$ is what the primary
calculated for the previous slot (even though that state is not
necessarily decided as of yet, and may never be decided).  Proposals
for a slot are stored in a set since a primary may propose to apply
different commands for the same slot due to repeated change of primaries.


Transition $\texttt{decide}(\slot, \cmd, \textit{res}, \textit{newState})$
specifies that only one of the proposed new states can be decided.
Because $\cmd$ was performed speculatively, the \texttt{decide}
transition checks that the state decided in the prior slot, if any,
matches state $s$ to which replica $r$ applied $\cmd$, thus
ensuring prefix ordering.

Similarly to active replication, transition \texttt{learn} models a replica
learning the decision of a slot.  With the \texttt{update} transition, a replica updates its state based on what was learned for the slot.
With active replication, each replica performs each client operation,
while in passive replication only the primary performs client operations
and backups simply obtain the resulting states.

Replicas wishing to act as primary perform transition \texttt{resetShadow} to update their speculative state and version, respectively denoted by variables $\shadowstate$ and $\shadowversion$.  The new shadow state may itself be speculative and must be set to a version at least as recent as the latest learned state.


To show that passive replication refines active replication, we first note that
all variables in passive replication have a one-to-one mapping with the ones in active replication, apart from
variables \textit{shadowVersion} and \textit{shadowState} which do not appear in active replication.  

Transition \texttt{resetShadow} correspond to a stutter transition.  Transitions \texttt{propose}, \texttt{decide}, and \texttt{learn} of passive
replication have preconditions that are either equal to the corresponding ones in active replication or more restrictive.  If these transitions are enabled in passive replication, they are therefore also enabled in active replication.  We now show that this is also the case for transition
 \texttt{update}.  For this to be true, (i) $\cmd$ must be different than $\bot$, and (ii) $(\textit{res}, \textit{newState}$) must be the result of applying command $\cmd$ on $\curstate$.  Condition (i) is trivially satisfied from Specification~\ref{alg:passive}.  Condition (ii) holds for the following reasons.  Since transition \texttt{update} is enabled, $\cmd$ is in $\learned[\version]$.  From transition \texttt{learn}, $\cmd$ was decided for slot $\version$.  From transition \texttt{decide}, $\cmd$ was proposed in slot $\version$ and either $\version = 1$ or $\version > 1$ and the state update \textit{cmd} was applied on the state decided in slot $\version~-~1$, that is, $\curstate$.  If $\version~=~1$, (\textit{res},~\textit{newState})~=~$\nextState(\state,~\cmd)$ trivially holds, otherwise it holds from the precondition of transition \texttt{propose}.

\begin{table*}
\begin{center}
{\footnotesize

\begin{tabular}{c||c|c|c|l}
Our term		& Paxos~\cite{L98}	& VSR~\cite{OL88}	& Zab~\cite{JRS11} & meaning		\\ \hline\hline
replica			& learner			& cohort		& server/observer & stores copy of application state	\\
certifier		& acceptor			& cohort		& server/participant & maintains consensus state \\
{\sequencer}	& leader			& primary		& leader & certifier that proposes orderings	\\
{\round}				& ballot			& view			& epoch	& round of certification \\
{\round}-id			& ballot number		& view-id		& epoch	number	& uniquely identifies a {\round} \\
normal case		& phase 2			& normal case	& normal case & processing in the absence of failures \\
recovery		& phase 1			& view change	& recovery & protocol to establish a new {\round}	\\
command			& proposal			& event record	& transaction & a pair of a client id and an operation to be performed \\
{\round}-stamp		& N/A				& viewstamp		& zxid & uniquely identifies a sequence of proposals
\end{tabular}
}
\end{center}
\caption{\label{tab:translate} {\small Translation between terms used in this paper
and in the various replication protocols under consideration.}}
\end{table*}

\mysection{A Generic Protocol}\label{sec:bevs}

Specifications~\ref{alg:active} and~\ref{alg:passive} contain
internal variables and transitions
that need to be refined for an executable implementation.
We start with refining active replication.
\multiconsensus\
(Specification~\ref{alg:bevs}) refines active replication and contains
no internal variables or transitions.
As previously, the \texttt{invoke} and \texttt{response} transitions (and corresponding variables) have been
omitted---they are the same as in Specification~\ref{alg:replsvc}. Transitions
\texttt{propose} and \texttt{update} of Specification~\ref{alg:active} have been omitted for the same reason.
In this section we explain how and why \multiconsensus\ works.


\mysubsection{Certifiers and {\round}s}

\multiconsensus\ has two basic building blocks:
\begin{itemize}
\item A static set of $n$ processes called \emph{certifiers}.
A minority of these may crash.  So for tolerating at most $f$
failures, we require that $n \ge 2f+1$ holds.
\item An unbounded number of \emph{{\round}}s.
\end{itemize}

\noindent
For ease of reference, Table~\ref{tab:translate}
contains a translation between terminology used in this paper
and those found in the papers describing the protocols under consideration.

In each round, a consensus protocol assigns to at most one certifier the role
of \emph{\sequencer}.
The {\sequencer} of a round can certify at most one command for each slot.
The other certifiers can copy the {\sequencer}, certifying the same
command for the same slot and {\round}.
Note that if two certifiers certify a command in the same
slot and the same {\round}, it must be the same command.
Moreover, a certifier cannot retract a certification.
Once a majority of certifiers certify
the command within a {\round}, the command is \emph{decided}
(and because certifications cannot be retracted the command will remain
decided thereafter).
In Section~\ref{sec:recovery} we show why
two {\round}s cannot decide different commands for the same slot.

Each {\round} has a {\round}-id that uniquely identifies the {\round}.
{\Round}s are totally ordered by their {\round}-ids.
A {\round} is in one of three modes: \emph{pending}, \emph{operational},
or \emph{wedged}.
One {\round} is the first {\round} (it has the smallest {\round}-id), and initially
only that {\round} is operational.  Other {\round}s start out pending.
The two possible transitions on the mode of a {\round} are as follows:
\begin{enumerate}
\item A pending {\round} can become operational only if all {\round}s with
lower {\round}-id are wedged;
\item A pending or operational {\round} can become wedged under any
circumstance.
\end{enumerate}

\noindent
This implies that at any time at most one {\round} is operational and that
wedged {\round}s can never become unwedged.

\mysubsection{Tracking Progress}\label{sec:progind}

In Specification~\ref{alg:bevs}, each certifier $\cert$ maintains a
\emph{progress indicator} $\progress[\slot]$ for each $\slot$, defined as:

\begin{definition}{Progress Indicator}
A progress indicator is a pair $\langle \bevId, \cmd \rangle$
where $\bevId$ is the identifier of a {\round} and $\cmd$ is a proposed command
or~$\bot$, satisfying:

\begin{itemize}
\item If $\cmd = \bot$, then the progress indicator guarantees that no {\round}
with an id less than $\bevId$ can ever decide, or have decided, a proposal
for the slot.
\item If $\cmd \ne \bot$, then the progress indicator guarantees that
if a {\round} with id $\bevId'$ such that $\bevId' ~\le~\bevId$ decides
(or has decided) a proposal $\cmd'$ for the slot, then $\cmd = \cmd'$.
\item Given two progress indicators $\langle \bevId, \cmd \rangle$ and
$\langle \bevId, \cmd' \rangle$ for the same slot, if neither $\cmd$ nor
$\cmd'$ equals $\bot$, then $\cmd = \cmd'$.
\end{itemize}
\end{definition}

\noindent
We define a total ordering $\succ$ on
progress indicators for the same slot as follows:
$\langle \bevId', \cmd' \rangle \succ \langle \bevId, \cmd \rangle$ iff
\begin{itemize}
\item $\bevId' > \bevId$; or
\item $\bevId' = \bevId \land \cmd' \ne \bot \land \cmd = \bot$.
\end{itemize}

\noindent
At any certifier, the progress indicator for a slot is monotonically
non-decreasing.


\begin{figure}[t]
\begin{center}

 \begin{algorithm}[H]

\caption{\label{alg:bevs} \multiconsensus}
\footnotesize{
\begin{distribalgo}[0]

\BLANK

\INDENT{\var ~ $\certBev, \primary, \progress[1...]$}
   \STATE $\certified, \snapshots$
\ENDINDENT

\BLANK

\INDENT{\init: $\certified = \snapshots = \emptyset ~\land$}
	\STATE $\forall \cert: \certBev = 0 \land \primary = \texttt{false} ~\land$
	\STATE $~~~~~~~~~~~~~~\forall \slot \in \nat^{+}: \progress[\slot] = \langle 0, \bot \rangle$
\ENDINDENT

\BLANK
\BLANK

\INDENT{\outtransition\ $\texttt{certifySeq}(\cert, \slot, \langle \bevId, \cmd \rangle)$:}
\INDENT{\precond:}
  \STATE $\primary \land \bevId = \certBev \land \progress[\slot] = \langle \bevId, \bot \rangle ~\land$
  \STATE $(\forall s \in \nat^+: \progress[s] = \langle \bevId, \bot \rangle \Rightarrow s \ge \slot) ~\land$
  \STATE $\exists \replica\ : \cmd \in \proposals[\slot]$
\ENDINDENT
\INDENT{\action:}
  \STATE $\progress[\slot] := \langle \bevId, \cmd \rangle$
  \STATE $\certified := \certified \cup \lbrace (\certId, \slot, \langle \bevId, \cmd\rangle) \rbrace$
\ENDINDENT
\ENDINDENT

\BLANK

\INDENT{\outtransition\ $\texttt{certify}(\cert, \slot, \langle \bevId, \cmd \rangle)$:}
\INDENT{\precond:}
  \STATE $\exists \cert': (\cert', \slot, \langle \bevId, \cmd \rangle) \in \certified ~\land$
  \STATE $\certBev = \bevId \land \langle \bevId, \cmd \rangle \succ \progress[\slot]$
\ENDINDENT
\INDENT{\action:}
  \STATE $\progress[\slot] := \langle \bevId, \cmd \rangle$
  \STATE $\certified := \certified \cup \lbrace (\certId, \slot, \langle \bevId, \cmd\rangle) \rbrace$
\ENDINDENT
\ENDINDENT

%
%
%
%

\BLANK

\INDENT{\outtransition\ $\texttt{\observedecision}(\replica, \slot, \cmd)$:}
\INDENT{\precond:}
  \STATE $\exists \bevId:$\\
       \hspace{0.05in} $ | \lbrace \cert ~|~ (\certId, \slot, \langle \bevId, \cmd \rangle) \in \certified \rbrace |  > \frac{n}{2} \land$
  \STATE $\learned[\slot] = \bot$
\ENDINDENT
\INDENT{\action:}
  \STATE $\learned[\slot] := \cmd$
\ENDINDENT
\ENDINDENT

\BLANK

\INDENT{\outtransition\ $\texttt{support{\Round}}(\cert, \bevId, \coord)$:}
\INDENT{\precond:}
  \STATE $\bevId > \certBev$
\ENDINDENT
\INDENT{\action:}
  \STATE $\certBev := \bevId; \primary := \texttt{false}$
  \STATE $\snapshots :=$\\
       \hspace{.08in} $ \snapshots \cup  \lbrace (\certId, \bevId, \coord, \progress) \rbrace$
\ENDINDENT
\ENDINDENT

\BLANK

\INDENT{\outtransition\ $\texttt{recover}(\cert, \bevId, {\cal S})$:}
\INDENT{\precond:}
  \STATE $\certBev = \bevId \land \lnot \primary \land |{\cal S}| > \frac{n}{2} ~\land$
  \STATE ${\cal S} \subseteq \lbrace (\mathit{id}, \progind) ~|~
	  (\mathit{id}, \bevId, \cert, \progind) \in \snapshots \rbrace$
\ENDINDENT
\INDENT{\action:}
  \INDENT{$\forall s \in \nat^{+}:$}
  	\STATE $\langle r, \cmd \rangle := \max_{\succ} \lbrace \progind[s] ~|~ (\mathit{id}, \progind) \in {\cal S} \rbrace$
	\STATE $\progress[s] := \langle \bevId, \cmd \rangle$
    \INDENT{\textbf{if} $\cmd \neq \bot$ \textbf{then}}
	  \STATE $\certified := \certified \cup \lbrace (\certId, s, \langle \certBev, \cmd\rangle) \rbrace$
    \ENDINDENT
  \ENDINDENT
  \STATE $\primary := \texttt{true}$
\ENDINDENT
\ENDINDENT

\end{distribalgo}
}
\end{algorithm}
\end{center}
\end{figure}

\floatname{algorithm}{Specification}
\renewcommand{\thealgorithm}{\arabic{algorithm}}

\mysubsection{Normal case processing}\label{sec:bev-normal}

Each certifier $\cert$ \emph{supports}
exactly one {\round}-id $\certBev$, initially~$0$, the {\round}-id of the
first {\round}.
The \emph{normal case} holds when a majority of certifiers support
the same {\round}-id, and one of these certifiers is {\sequencer} (its $\primary$ flag is set to \texttt{true}).

Transition $\texttt{certifySeq}(\cert, \slot, \langle \bevId, \cmd \rangle)$
is performed when {\sequencer} $\cert$ certifies command $cmd$ for the given slot
and {\round}.
The condition $\progress[\slot] = \langle \bevId, \bot \rangle$
holds only if no command can be decided in this slot
by a {\round} with an id lower than $\certBev$.
The transition requires that $\slot$ is the lowest empty slot of the
sequencer.
If the transition is performed,
$\cert$ updates $\progress[\slot]$ to reflect that if a command is
decided in its {\round}, then it must be command $\cmd$.
{\Sequencer} $\cert$ also notifies all other certifiers by adding
$(\certId, \slot, \langle \bevId, \cmd \rangle)$ to set $\certified$
(modeling a broadcast to the certifiers).

A certifier that receives such a message checks if the message
contains the same {\round}-id that it is currently supporting and that
the progress indicator in the message exceeds its own progress
indicator for the same slot.  If so, then the certifier updates its
own progress indicator and certifies the proposed command (transition
$\texttt{certify}(\cert, \slot, \langle \bevId, \cmd \rangle)$).

The $\texttt{\observedecision}(\replica, \slot, \cmd)$
transition at $\replica$ is enabled if a majority of
certifiers in the same {\round} have certified $\cmd$ in $\slot$.
If so, the command is decided and, as explained in the next section,
all replicas that undergo the \observedecision\ transition
for this slot will decide on the same command.

\mysubsection{Recovery}\label{sec:recovery}

In this section, we show how \multiconsensus\ deals with failures.
The reason for having an unbounded number of {\round}s is to achieve liveness.
When an operational {\round} is no longer certifying proposals,
perhaps because its {\sequencer} has crashed or is slow,
the round can be
wedged and a {\round} with a higher {\round}-id can become operational.

Modes of {\round}s are implemented as follows:
A certifier $\cert$ can transition to supporting a new {\round}-id
$\bevId$ and \emph{prospective sequencer} $\coord$
(transition $\texttt{support{\Round}}(\cert, \bevId, \coord)$).
This transition can only increase $\certBev$.
Precondition $\bevId~>~\certBev$ ensures that a certifier supports a {\round} and a prospective sequencer for the round at most once.
The transition sends the certifier's \emph{snapshot} by adding it to the
set $\snapshots$.
A snapshot is a four-tuple $(\certId, \bevId, \coord, \progress)$ containing
the certifier's identifier, its current {\round}-id, the identifier of $\coord$, and
the certifier's list of progress indicators.
Note that a certifier can send at most one snapshot for each {\round}.

{\Round} $\bevId$ with sequencer $\coord$ is operational,
by definition, if a majority of certifiers support $\bevId$ and added
$(\certId, \bevId, \coord, \progress)$ to the set \textit{snapshots}.
Clearly, the majority requirement guarantees that there cannot be two
{\round}s that are simultaneously operational, nor can there be operational
{\round}s that do not have exactly one sequencer.
Certifiers that support $\bevId$ can no longer certify commands
in {\round}s prior to $\bevId$.
Consequently, if a majority of certifiers support a {\round}-id
larger than $x$, then all {\round}s with an id of $x$ or lower
are wedged.

\begin{figure*}
\begin{center}
\begin{math}
\begin{array}{rcl}
{\decisions[\slot]} & = & {\left\{
	\begin{array}{cl}
		\cmd & {\textbf{if}~~ \exists \bevId:
			| \lbrace \cert ~|~ (\certId, \slot, \langle \bevId, \cmd \rangle) \in \certified \rbrace | > n/2} \\
		\bot & \textbf{otherwise}
	\end{array}
\right.}
\end{array}
\end{math}
\end{center}

\caption{\label{fig:refine} {\small Relation between the $\certified$
variable of Specification~\ref{alg:bevs} and the
$\decisions$ variable of Specification~\ref{alg:active}.
Here $n$ is the number of certifiers.}}

\end{figure*}

Transition $\texttt{recover}(\cert, \bevId, {\cal S})$
is enabled at $\cert$ if
the set $\cal S$ contains snapshots for $\bevId$ and sequencer $\cert$
from a majority of certifiers.
The sequencer
helps to ensure that the {\round} does not
decide commands inconsistent with prior {\round}s using the snapshots
it has collected.
For each slot, sequencer $\cert$ determines the maximum
progress indicator $\langle r, \cmd \rangle$ for the slot in the
snapshots contained in $\cal S$.  It then sets its own progress
indicator for the slot to $\langle \bevId, \cmd \rangle$.
It is easy to see that $\bevId \geq r$.

We argue that $\langle \bevId, \cmd \rangle$ satisfies the definition
of progress indicator in Section~\ref{sec:progind}.  All certifiers
in $\cal S$ support $\bevId$ and form a majority.  Thus, it is not
possible for any {\round} between $r$ and $\bevId$ to decide a command
because none of these certifiers can certify a command in those
{\round}s.  There are two cases:

\begin{itemize}
\item If $\cmd = \bot$, no command can be
decided before $r$, so no command can be decided before $\bevId$.
Hence, $\langle \bevId, \bot \rangle$ is a correct progress indicator.
\item If $\cmd \ne \bot$, then if a command is decided by
$r$ or a {\round} prior to $r$, it must be $\cmd$.
Since no command can be decided by {\round}s between $r$ and $\bevId$,
$\langle \bevId, \cmd \rangle$ is a correct progress indicator.
\end{itemize}


\noindent
The sequencer sets its $\primary$ flag upon recovery.  As a result, it
is enabled to propose new commands. Normal case for the {\round} begins
and holds as long as a majority of certifiers support the corresponding
{\round}-id.

\mysubsection{Refinement Mapping}

\multiconsensus\ refines active replication.
We first show how the internal variables of Specification~\ref{alg:active} are
derived from the variables in Specification~\ref{alg:bevs}.
Predicate $\decisions[\slot] = \cmd$ holds if there exists any {\round} that has decided
the command (Section~\ref{sec:recovery} argues why all {\round}s of a given slot can only decide the same command); otherwise $\decisions[\slot] = \bot$.
This is captured formally in
Fig.~\ref{fig:refine}, where $\cmd$ is a tuple $(\client, \op)$.

The transition $\texttt{certifySeq}(\cert, \slot, \langle \bevId, \cmd \rangle)$
 of \multiconsensus\ always corresponds to a stutter in active replication.
The $\texttt{decide}(\slot, \cmd)$ transition
of Specification~\ref{alg:active} is performed
when, for the first time, a majority of certifiers in some
{\round} $\bevId$ have certified command $\cmd$
in slot $\slot$, that is, when
the last certifier $\cert$ in the majority performs transition
$\texttt{certify}(\cert, \slot, \langle \bevId, \cmd \rangle)$.
The \texttt{\observedecision} transition of \multiconsensus\ corresponds exactly
to the \texttt{learn} transition of active replication.  Both the \texttt{support{\Round}} and \texttt{recover} transitions are
stutters with respect to Specification~\ref{alg:active} as they
do not affect any of its state variables.

\mysubsection{Passive Replication}\label{sec:bev-prefix}

Section~\ref{sec:ap-formal} showed how in passive replication a state
update from a particular primary can only be decided in a slot if it corresponds to applying an operation
on the state decided in the previous slot.
We called this property prefix ordering.
However, Specification~\ref{alg:bevs} does not satisfy prefix ordering
because any proposal can be decided in a slot, in particular one that
does not correspond to a state decided in the prior slot.
Thus \multiconsensus\ does not refine Passive Replication.
One way of implementing prefix ordering would be for the primary to
wait with proposing a command for a slot until it knows the decisions
for all prior slots.
Doing so would be slow.

A better solution is to refine \multiconsensus\ to obtain a
specification that also refines Passive Replication and satisfies
prefix ordering.  We call this specification \multiconsensus-PO.
\multiconsensus-PO guarantees that each decision is the result
of an operation applied to the state decided in the prior slot
(except for the first slot).
We complete the refinement by adding two preconditions:
\begin{enumerate}
\item[(i)]
In \multiconsensus-PO, slots have to be decided in order.
To guarantee this, we have each certifier certify commands
in order by adding the following
precondition to transition \texttt{certify}:
$\slot > 1 \Rightarrow$ $\exists c \ne \bot: \progress[\slot - 1] = \langle \bevId, c \rangle$.
Thus if, in some round, there exists a majority of certifiers that have
certified a command in $\slot$, there also exists a majority of certifiers
that have certified a command in the prior slot.

\item[(ii)]
To guarantee that a decision in $\slot$ is based on the state
decided in the prior slot, we add the following precondition to transition \texttt{certifySeq}:
$\slot > 1 \Rightarrow \exists~\textit{oldState}~:~\cmd~=~(\textit{oldState}, -) \land \progress[\slot~-~1]~=~\langle \bevId, (-, (-,-,\textit{oldState})) \rangle$.
This works because by the properties of progress indicators, if a
command has been or will be decided in round $\bevId$ or a prior round,
it is the command in $\progress[\slot-1]$.
Therefore, if the sequencer's proposal for $\slot$ is decided in round $\bevId$, it is the command in $\progress[\slot-1]$.
If the primary and the sequencer are co-located, as they usually are,
this condition is satisfied automatically as the primary computes
states in order.
\end{enumerate}

Also, \multiconsensus-PO inherits transitions \texttt{invoke} and \texttt{response} from Specification~\ref{alg:replsvc} as well as transitions
\texttt{propose}, \texttt{update}, and \texttt{resetShadow} from Specification~\ref{alg:passive}.  The variables contained in these transitions are inherited as well.

The passive replication protocols that we consider in this paper,
VSR and Zab,
share the following design decision in the recovery procedure: The
{\sequencer} broadcasts a single message containing its entire
snapshot rather than sending separate certifications for each slot.
Certifiers wait for this comprehensive snapshot, and overwrite their
own snapshot with it, before they certify new commands in this {\round}.
As a result, at a certifier all $\progress$ slots have the same {\round}
identifier, and can thus be maintained as a separate variable.  

\subsubsection{Refinement Mappings}
\label{sec:MCPO:refinement}

Below, we present a refinement between \multiconsensus-PO\ and \multiconsensus\ and then show that \multiconsensus-PO refines Passive Replication.

\subsubsection{Refining \multiconsensus}

Showing the existence of a refinement between \multiconsensus-PO\
and \multiconsensus\ is straightforward.  Transitions and variables
inherited from passive replication are mapped to variables and
transitions of \multiconsensus\ in the same way as they are mapped
from passive replication to active replication (note that these
variables and transitions are mapped to variables and transitions
of \multiconsensus\ that are themselves inherited from active
replication).  Since \multiconsensus-PO\ only adds constraints to
the transitions that are specific to \multiconsensus, the refinement
exists.

\subsubsection{Refining Passive Replication}
Passive replication has a single internal variable, $\decisions$,
that is derived from $\certified$ as in Fig.~\ref{fig:refine},
that is, for any slot $s$, $\decisions[s] = \cmd$ holds if a majority
of replicas have certified $\cmd$ for slot $s$.

Transitions \texttt{certifySeq}, \texttt{support{\Round}}, and
\texttt{recover} are stutters in passive replication, while an
\texttt{\observedecision} transition of \multiconsensus-PO
corresponds to a \texttt{learn} transition in
Specification~\ref{alg:passive}.

Similarly to the refinement between active replication and
\multiconsensus, transition $\texttt{decide}(\slot, \cmd, \textit{res},
\textit{newState})$ is performed when, for the first
time, a majority of certifiers in some {\round} $\bevId$ have certified
command $\cmd$ in slot $\slot$, that is, when the last certifier
$\cert$ in the majority performs transition $\texttt{certify}(\cert,
\slot, \langle \bevId, \cmd \rangle)$.  

In this case however, we
must additionally show that when the last certifier of the majority
undergoes the \texttt{certify} transition, the following condition
holds in Specification~\ref{alg:passive}: $\exists r, \textit{oldState} : (\textit{oldState},
(\cmd, \textit{res}, \textit{newState})) \in \textit{proposals}_r[\slot]$
and $(\slot > 1 \Rightarrow \decisions[\slot-1] = (-,-,\textit{oldState}))$.  

The first
part of the condition holds since certifiers only certify commands
that have been proposed.  The second part of the condition holds
for the following reason.  From constraint (ii) of \multiconsensus-PO, the sequencer only certifies
a command if its corresponding \textit{oldState} equals the \textit{newState} of the command it stores in the previous slot.
From the properties of progress indicators, if $\progress[\slot-1] =  \langle \bevId, \cmd \rangle$, then if a command is decided
in an earlier round than $\bevId$, then it must be $\cmd$.  From constraint (i) (commands are certified in order), if a command is decided for $\slot$, then
$\slot - 1$ decided on a command previously.  Consequently,  $(\slot > 1 \Rightarrow \decisions[\slot-1] = (-,-,\textit{oldState}))$.

\mysection{Implementation}\label{sec:implementation}

Specifications \multiconsensus\ and \multiconsensus-PO do not contain
internal variables or transitions.  However, they only specify which
transitions are safe, not which transitions to perform and at what
time.  We show, informally, final refinements of
these specifications to obtain Paxos, VSR, and Zab.

\mysubsection{Normal Case}\label{sec:impl-normal-case}

We first turn to implementing the state and transitions of \multiconsensus\ and \multiconsensus-PO.
The first question is how to implement the variables.
Variables $\inputs$, $\outputs$, $\certified$, and $\snapshots$ are
not per-process but global.  They model messages that have been sent.
In actual protocols, these are implemented
by the network: a value in either set is implemented by a message
on the network tagged with the appropriate type, such as
\texttt{snapshot}.


The remaining variables are all local to a process such as a client,
a replica, or a certifier, and can be implemented as ordinary
program variables.
In Zab, the $\progress$ variable is implemented by a queue
of commands.
In VSR, the $\progress$ variable is replaced by the application state
and a counter that counts the number of updates made to the state in
this {\round}.
In Paxos, a progress indicator is simply a pair consisting of a
round identifier and a command.

\begin{figure}
  \centering
	\includegraphics[scale=.7]{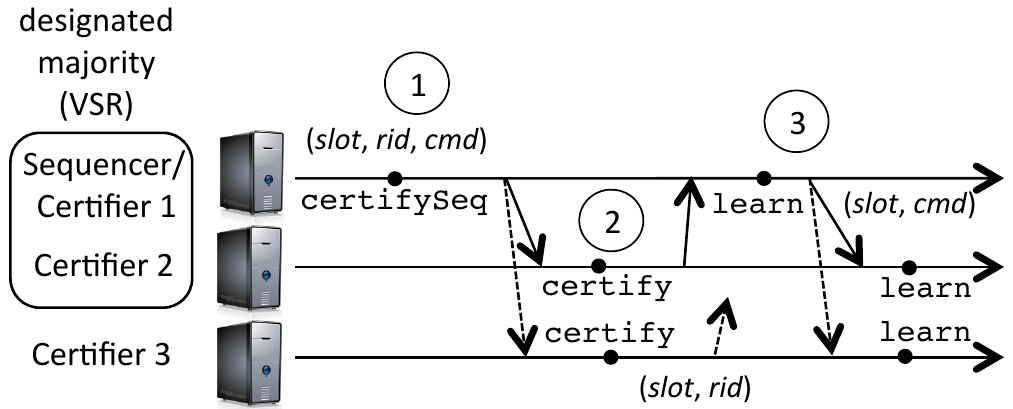}
  \caption{\label{fig:protocol_normal_case} {\small Normal case
processing at three certifiers.
Dots indicate transitions, and arrows between certifiers are messages.}}
\end{figure}

\begin{figure*}
  \centering
  \begin{tabular}{@{\hspace{-.1cm}}c @{\hspace{-.1cm}} c@{\hspace{-.1cm}} c@{\hspace{0cm}}}
  \subfloat[Paxos]{
    \label{fig:paxos-recovery}
    \includegraphics[scale=.62]{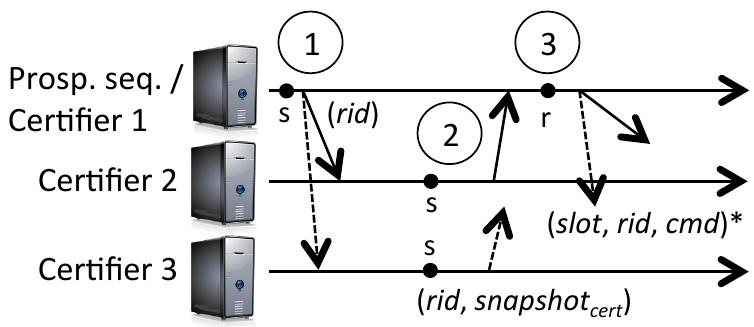}
  }
&
  \subfloat[Zab]{
    \label{fig:zab-recovery}
    \includegraphics[scale=.62]{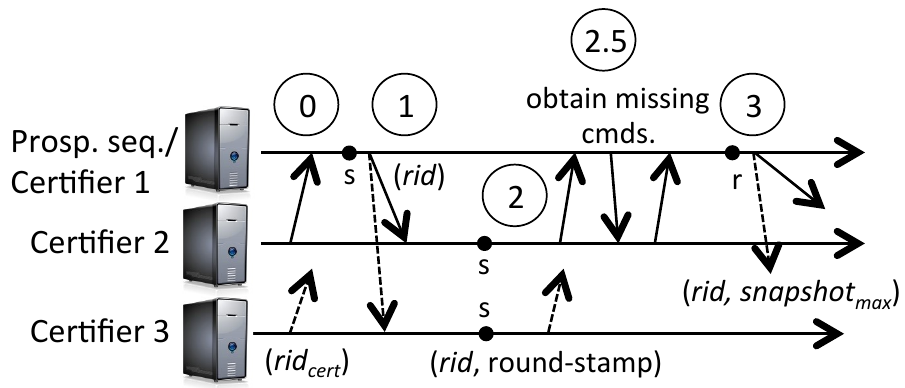}
  }
  &
  \subfloat[VSR]{
    \label{fig:vsr-harp-recovery}
    \includegraphics[scale=.62]{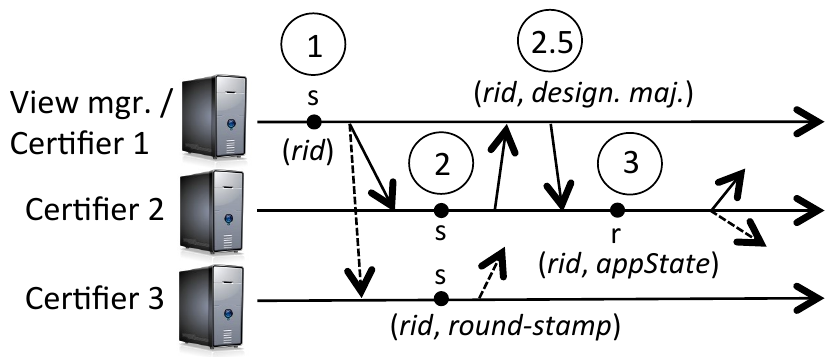}
  }
  \end{tabular}
  \caption{{\small Depiction of the recovery phase. Dots represent transitions and are labeled with \texttt{s} and \texttt{r}, respectively denoting a \texttt{support{\Round}} and a \texttt{recover} transition. Messages of the form (x, y)* contain multiple (x,y) tuples.}}
  \label{fig:protocol_recovery}
\end{figure*}

Fig.~\ref{fig:protocol_normal_case} illustrates the steps of
normal case processing in the protocols.
The figure shows three certifiers ($f = 1$).
Upon receiving an operation from a client (not shown):
\begin{enumerate}
\item
The {\sequencer} $\cert$ proposes a command for the next open slot
and sends a message to the other certifiers (maps to
$\texttt{certifySeq}(\cert, \slot, \langle \bevId, \cmd \rangle)$).  In the case of VSR
and Zab, the command is a state update that results from executing
the client operation; with Paxos, the command is the
operation itself.

\item Upon receipt by a certifier $\cert$, if $\cert$ supports the {\round}-id $\bevId$
in the message, then $\cert$ updates its slot and replies to the {\sequencer}
(transition $\texttt{certify}(\cert, \slot, \langle \bevId, \cmd \rangle)$).  With VSR and Zab, prefix-ordering
must be ensured, and $\cert$ only replies to the \sequencer\ if its progress indicator for $\slot-1$ contains a non-empty command for $\bevId$.

\item If the {\sequencer} receives successful responses from a majority of
certifiers (transition \observedecision), then the {\sequencer} learns the decision and broadcasts
a \texttt{decide} message for the command to the replicas
(resulting in \texttt{learn} transitions that update the replicas (see Specifications~\ref{alg:active} and~\ref{alg:passive}).
\end{enumerate}

\noindent
The various protocols make additional design decisions:
\begin{itemize}
\item
In the case of VSR, a specific majority of certifiers
is determined a priori and fixed for each {\round}.
We call this a \emph{designated majority}.
In Paxos and Zab, any certifier can certify proposals.

\item
In VSR, replicas are co-located with certifiers, and certifiers
speculatively update their local replica as part of certification.
A replica may well be updated before some
proposed command is decided, so if another command is decided
the state of the replica must be rolled back, as we shall see later.
Upon learning that the command has been decided (Step~3), the {\sequencer}
responds to the client.

\item
Optionally, Paxos uses leases~\cite{GC89,L98} for read-only operations.
Leases have the advantage that read-only operations can be served at a single replica while still 
guaranteeing linearizability. This method assumes synchronized clocks (or clocks with bounded drift) and has the {\sequencer} obtain a lease for a certain time period.  A {\sequencer} that is holding a lease can thus forward read-only operations
to any replica, inserting the operation in the ordered stream of commands
sent to that replica.

\item
Zab (or rather ZooKeeper) offers the option to use leasing or to have
any replica handle read-only operations
individually, circumventing Zab.
The latter is efficient, but a replica may not have learned the latest
decided proposals and its clients receive results based on
stale state (such reads satisfy sequential consistency~\cite{Lamport79}).
\end{itemize}

\noindent
For replicas to learn about decisions, two options exist:
\begin{itemize}
\item Certifiers can respond back to the {\sequencer}.  The {\sequencer} learns
that its proposed command has been decided if the {\sequencer} receives
responses from
a majority (counting itself).  The {\sequencer} then notifies the replicas.
\item Certifiers can broadcast notifications to all replicas,
and each replica can individually determine if a
majority of the certifiers have certified a particular command.
\end{itemize}

\noindent
There is a trade-off between the two options:
with $n$ certifiers and $m$ replicas,
the first approach requires $n + m$ messages and two network latencies.
The second approach requires $n \times m$ messages but involves only one
network latency.
All implementations we know of use the first approach.




\mysubsection{Recovery}\label{sec:bev-rec-real}

Fig.~\ref{fig:protocol_recovery} illustrates the recovery steps in
the protocols.


With Paxos, a certifier $\coord$ unhappy with progress starts the following
process to try to become sequencer itself
(see Fig.~\ref{fig:paxos-recovery}):
\begin{itemize}
\item Step 1: Prospective sequencer $\coord$ supports a new {\round} $\bevId$ proposing
itself as sequencer
(transition $\texttt{support{\Round}}(\coord, \bevId, \coord)$), and
queries at least a majority of certifiers.
\item Step 2: Upon receipt, a certifier $\cert$
that transitions to supporting {\round} $\bevId$ and certifier~$\coord$
as sequencer
($\texttt{support{\Round}}(\cert, \bevId, \coord)$) responds with its snapshot.
\item Step 3: Upon receiving responses from a majority,
certifier~$\coord$
learns that it is sequencer of {\round} $\bevId$ (transition $\texttt{recover}(\coord, \bevId, {\cal S})$).  Normal case operation resumes after
$\coord$ broadcasts the command with the highest
$\bevId$ for each slot not known to be decided.
\end{itemize}

\vspace{.1in}

Prospective sequencers in Zab are determined by a weak leader election protocol $\Omega$~\cite{ST08}.
When $\Omega$ determines that a sequencer has become unresponsive, it
initiates a protocol (see Fig.~\ref{fig:zab-recovery})
in which a new sequencer is elected:
\begin{itemize}
\item Step 0: $\Omega$ proposes a prospective sequencer $\coord$ and notifies
the certifiers.
Upon receipt, a certifier sends a message containing the {\round}-id it supports to $\coord$.
\item Step 1: Upon receiving such messages from a majority of certifiers,
prospective sequencer $\coord$ selects a {\round}-id $\bevId$ that is one larger
than the maximum it received, transitions to supporting it
(transition $\texttt{support{\Round}}(\coord, \bevId, \coord)$),
and broadcasts this to the other certifiers for approval.
\item Step 2: Upon receipt, if
certifier $\cert$ can support $\bevId$ and has not agreed to a certifier
other than $\coord$ becoming sequencer of the {\round}, it performs
transition $\texttt{support{\Round}}(\cert, \bevId, \coord)$.
Zab exploits prefix ordering to optimize the recovery protocol.
Instead of sending its entire snapshot to the prospective sequencer,
a certifier that transitions to supporting {\round} $\bevId$ sends a \emph{{\round}-stamp}.
A {\round}-stamp is a lexicographically ordered pair consisting of the
{\round}-id in the snapshot (the same for all slots) and the number of slots
in the {\round} for which it has certified a command.
\item Step 2.5: 
If $\coord$ receives responses from a majority, $\coord$ computes the maximum {\round}-stamp and determines
if it is missing commands. If it is the case, $\coord$ retrieves them from the certifier $\cert_{max}$ with the highest received
{\round}-stamp.  If $\coord$ is missing too many commands (e.g. if $\coord$ did not participate in the last {\round} $\cert_{max}$ participated in), $\cert_{max}$ sends its
entire snapshot to $\coord$.  
\item Step 3:
After receiving the missing commands, $\coord$ broadcasts its snapshot, in practice a checkpoint of its state with a sequence of state updates, to the certifiers
(transition $\texttt{recover}(\coord, \bevId, {\cal S})$).  Certifiers acknowledge the reception of
this snapshot and, upon receiving acknowledgments from a majority, $\coord$ learns that it is now the \sequencer\ of $\bevId$ and broadcasts a \emph{commit} message before resuming
 the normal case protocol (not shown in the picture).
\end{itemize}



In VSR, each {\round}-id has a pre-assigned \emph{view manager} $v$ that is not necessarily the sequencer.  
A {\round}-id is a
lexicographically ordered pair comprising a number and the process identifier
of the view manager.

If unhappy with progress, the view manager $v$ of {\round}-id starts the following
recovery procedure (see Fig.~\ref{fig:vsr-harp-recovery}):
\begin{itemize}
\item Step 1: $v$ starts supporting {\round} $\bevId$
(transition $\texttt{support{\Round}}(v, \bevId, v)$), and
queries at least a majority of certifiers.
\item Step 2:
Upon receipt of such a query, a certifier $\cert$ starts supporting
$\bevId$ (transition $\texttt{support{\Round}}(\cert, \bevId, v)$).
Similarly to Zab, $\cert$ sends its {\round}-stamp to $v$.
\item Step 2.5:
Upon receiving {\round}-stamps from a majority of certifiers,
view manager $v$ uses the set of certifiers that responded
as the designated majority for the {\round} and assigns the {\sequencer} role
to the certifier $p$ that reported the highest {\round}-stamp.  The view manager
then notifies certifier $p$, requesting it to become {\sequencer}.
\item Step 3:
Sequencer $p$, having the latest state, broadcasts its snapshot
(transition $\texttt{recover}(p, \bevId, {\cal S})$).
In the case of VSR, the state that the new {\sequencer} sends is
its application state rather than a snapshot. 
\end{itemize}


\mysubsection{Garbage Collection}\label{sec:bev-rec}

\multiconsensus\ has each certifier building up state about
all slots, which does not scale.
Unfortunately, little is written about this issue in the Paxos
and Zab papers.
%
%
In VSR, no garbage collection is required.  Certifiers and replicas are
co-located, and only store the most recent
{\round}-id they adopted and the application state that is updated upon
certification of a command.
During recovery the {\sequencer} simply sends
the application state to the replicas, and consequently, there is no need to
replay any decided commands.

\mysubsection{Liveness}\label{sec:liveness}

All of the protocols require---in order to make progress---that at
most a minority of certifiers experience crash failures.  If the current
{\round} is no longer making progress, a new {\round} must become operational.  If certifiers are slow at this
in the face of an actual failure, then performance may suffer.
However, if certifiers are too aggressive, {\round}s will become wedged
before being able to decide commands, even in the absence of failures.

To guarantee progress, some {\round} with a correct {\sequencer}
must eventually not get preempted by a higher {\round}~\cite{DLS84old,CT91}.  Such a guarantee is difficult or even impossible to
make~\cite{FLP85}, but with careful failure detection a good trade-off
can be achieved between rapid failure recovery and
spurious wedging of {\round}s~\cite{KA08}.

In this section, we will look at how the various protocols try optimizing
progress.

\mysubsubsection{Partial Memory Loss}\label{sec:memloss}

If certifiers keep their state on stable storage (say, a disk),
then a crash followed by a recovery is not treated as a failure but
instead as the affected certifier being slow.  Stable storage allows
protocols like Paxos, VSR, and Zab to deal with such transients.
Even if all machines crash, as long as a majority eventually recovers
their state from before the crash, the service can continue operating.

\mysubsubsection{Total Memory Loss}\label{sec:totalloss}

In Section 5.1 of ``Paxos Made Live''~\cite{CGR07}, the
developers of Google's Chubby service describe
a way for Paxos to deal with permanent memory loss of a certifier
(due to disk corruption).
The memory loss is total, so the recovering certifier starts in an
initial state.  It copies its state from another certifier and then
waits until it has seen one decision before starting to participate
fully in the Paxos protocol again.  This optimization is flawed since it breaks the invariant that
a certifier's {\round}-id can only increase over time (confirmed by the authors of~\cite{CGR07}).  By copying the state
from another certifier, it may, as it were, go back in time, which
can cause divergence.


Nonetheless, total memory loss can be tolerated by extending the protocols.
The original Paxos paper~\cite{L98} shows how the set of certifiers
can be reconfigured to tolerate total memory loss, and this has been worked out in greater detail
in Microsoft's SMART project~\cite{LABCDH06} and later for Viewstamped
Replication as well~\cite{liskov12vr}.
Zab also supports reconfiguration~\cite{SRMJ12}.

\begin{table*}
\begin{center}
{\footnotesize

\begin{tabular}{c|c||c|c|c}
What					& Section					& Paxos		& VSR			& Zab		\\ \hline\hline
replication style		& \ref{sec:actpass}			& active	& passive		& passive	\\
read-only operations	& \ref{sec:impl-normal-case} & leasing	& certification	& read any replica/leasing \\
designated majority		& \ref{sec:impl-normal-case}, \ref{sec:bev-rec-real} & no			& yes & no	\\
time of execution		& \ref{sec:impl-normal-case} & upon decision		& upon certification & depends on role \\
{\sequencer} selection	& \ref{sec:bev-rec-real}	& majority vote	or deterministic & view manager assigned & majority vote \\
recovery direction		& \ref{sec:bev-rec-real}	& two-way	& from {\sequencer} & two-way/from sequencer \\
recovery granularity	& \ref{sec:bev-rec-real}	& slot-at-a-time	& application state	& command prefix \\
tolerates memory loss	& \ref{sec:memloss}, \ref{sec:totalloss} & reconfigure & partial		& reconfigure
\end{tabular}
}
\end{center}
\caption{\label{tab:diffs} {\small Overview of important
differences between the various protocols.}}
\end{table*}

\mysection{Discussion}\label{sec:discuss}

Table~\ref{tab:diffs} summarizes differences between Paxos, VSR,
and Zab.  We believe that these differences are important
because they both demonstrate that the protocols do not refine one
another, and the differences have pragmatic
consequences as discussed below.  The comparisons are based on
published algorithms; actual implementations may vary.  We organize the
discussion around normal case processing and recovery overheads.


\mysubsection{Normal Case}

\myparagraph{Passive vs. Active Replication}

In active replication, there are at least $f+1$ replicas that each
have to execute operations.  In passive replication,
only the {\sequencer} executes operations, but has to propagate state updates
to the backups.  Depending on the overheads
of executing operations and the size of state update messages, one
or the other may perform better.  Passive replication has the
advantage that execution at the {\sequencer} does not have to be
deterministic and can take advantage of parallel processing on
multiple cores.  

\myparagraph{Read-only Optimizations}

Paxos and Zookeeper support leasing for read-only operations,
but there is no reason why leasing could not be added to VSR as well.
Indeed, Harp (built on VSR) uses leasing.
A lease improves latency of read-only operations in the normal
case, but delays recovery in case of a failure.
ZooKeeper offers the option whether leases should be used.
Without leases, clients read any replica at any time.
Doing so compromises consistency since replicas
may have stale state.

\myparagraph{Designated Majority}

VSR uses designated majorities.  This has the advantage that the
other (typically~$f$) certifiers and replicas are not employed during normal
operation, and they play only a small role during recovery, saving almost
half of overhead.  
There are two disadvantages:
(1) if the designated majority contains the slowest certifier the protocol will run at the rate of that slowest certifier, as opposed to the ``median'' certifier; and
(2) if one of the certifiers in
the designated majority crashes or becomes unresponsive or slow, then a
recovery is necessary.
In Paxos and Zab, recovery is necessary only if the {\sequencer} crashes.
A middle ground can be achieved by using $2f+1$ certifiers and $f+1$
replicas.

\myparagraph{Time of Command Execution}

In VSR, replicas apply state updates speculatively at the same time that
they are certified, possibly before they are decided.
Commands are forgotten as soon as they are applied to the state.
This means that no garbage collection is necessary.  A disadvantage
is that the response to a client operation must be delayed until
all replicas in the designated majority have updated their application
state.  In other protocols, only one replica has to have updated
its state and computed a response, because in case the replica fails
another deterministic replica is guaranteed to compute the same
response.  Note that at this time each command that led to this
state and response has been certified by a majority and therefore
the state and response are recoverable even if this one replica
crashes.

In Zab, the primary also speculatively applies client operations
to compute state updates before they are decided.
However, replicas only apply those state updates until after they have
been decided.
In Paxos, replicas only execute a command
after it has been decided and there is no speculative execution.


\mysubsection{Recovery}


\myparagraph{{\Sequencer} Selection}


VSR provide an advantage in selecting a
{\sequencer} that has the most up-to-date state (taking advantage of prefix
ordering), and thus it does not have to recover this
state from the other certifiers, simplifying and streamlining recovery.

\myparagraph{Recovery Direction}

Paxos allows the prospective {\sequencer} to recover the state
of previous slots and, at the same time, propose new commands for
slots for which it already has retrieved sufficient state.  
However, all certifiers must
send their certification state to the prospective {\sequencer} before it re-proposes commands for slots.  With VSR, the sequencer is the certifier
with the highest {\round}-stamp and it does not need to recover state from the other certifiers.  A similar optimization
is sketched in the description of the Zab protocol (and implemented in the Zookeeper service).


With Paxos and Zab, garbage collection of the certification state is important to ensure that the amount of state that has to be exchanged on
recovery does not become too large.
It is often faster to bring
a recovering replica up to date by replaying decided commands that it missed
rather than by copying state.  

With VSR, the selected {\sequencer} pushes its snapshot to the other certifiers.  The snapshot has to be transferred and processed
before new certification requests, possibly resulting in a performance hiccup.

\myparagraph{Recovery Granularity}

In VSR, state sent from the {\sequencer} to the backups is the entire
application state.  For VSR, this state is transaction manager state
and is small, but in general such an approach does not scale.
However, in some cases that cost is unavoidable, even in the other
protocols.  For example, if a replica has a disk failure, replaying
all commands from day~0 is not scalable either, and the recovering
replica instead will have to seed its state from another one.  In
such a case, the replica will load a checkpoint and then replay
missing commands to bring the checkpoint up-to-date---this technique
is used in Zab (and in Harp as well).  With passive replication protocols,
replaying missing commands simply means applying state updates;
with active replication protocols, replaying commands entails
re-executing commands.  Depending on the
overheads of executing operations and the size of state update
messages, one or the other approach may perform better.

\myparagraph{Tolerating Memory Loss}

An option suggested by VSR is to keep only a {\round}-id
on disk; the remaining of the state is in memory.  This technique
works only in restricted situations where at least one certifier
has the most up-to-date state in memory.

\mysection{A Bit of History}\label{sec:history}

\noindent
Based on the discussion so far one may think that {\round}s and sequencers 
were first introduced by protocols that implement \multiconsensus.
However, we believe the first consensus protocol to use
{\round}s and sequencers is due to
Dwork, Lynch, and Stockmeyer (DLS)~\cite{DLS84old}.
Rounds in DLS are countable and {\round} $b+1$ cannot start until
{\round} $b$ has run its course.
Thus, DLS does not refine \multiconsensus.

Chandra and Toueg's work on consensus~\cite{CT91} formalized the conditions under which consensus protocols terminate by encapsulating synchrony assumptions in the form of failure detectors.  Their consensus protocol resembles the Paxos single-decree Synod protocol and refines \multiconsensus.


To the best of our knowledge,
the idea of using majority intersection to avoid potential inconsistencies
first appears in Thomas~\cite{Tho78}.
Quorum replication~\cite{Tho78} 
supports only storage objects with \texttt{read} and \texttt{write}
operations (or, equivalently, \texttt{get} and \texttt{put} operations in the
case of a Key-Value Store).

\mysection{Conclusion}\label{sec:conclusion}

Paxos, VSR, and Zab are well-known replication protocols for
an asynchronous environment that admits a bounded number of crash
failures.
The paper describes a specification for \multiconsensus, a generic
specification that contains important design features that the protocols share.
These features include an unbounded number of totally ordered {\round}s,
a static set of certifiers, and at most one {\sequencer} per {\round}.

The protocols differ in how they refine \multiconsensus.
We were able to disentangle fundamentally
different design decisions in the three protocols and consider their
impact on performance.
Most importantly, compute-intensive services are better off with a
passive replication strategy such as used in VSR and Zab
(provided that state updates are of a reasonable size). 
To achieve predictable low-delay performance for short operations during
both normal case execution and recovery, an active replication strategy
without designated majorities, such as used in Paxos, is the best option.

\begin{small}

\end{small}

\end{document}